\documentclass[12pt]{article}
\pdfoutput=1
\usepackage{amsmath,amssymb,amsthm,bm,graphicx,float,array,multirow,multicol,rotfloat,caption,subcaption,hyperref,cleveref,enumerate,geometry,mathdots,adjustbox,booktabs,parskip,mathtools,tikz,tikz-cd,pdflscape,csquotes,lscape,rotating,empheq,mathrsfs,longtable}
\usepackage[para]{threeparttable}
\usepackage[all]{xy}
\usepackage[normalem]{ulem}
\usepackage[aligntableaux=center]{ytableau}
\usepackage[numbers,sort&compress]{natbib}
\numberwithin{equation}{section}
\numberwithin{figure}{section}
\allowdisplaybreaks
\makeatletter

\usepackage{comment}

\theoremstyle{plain}
\newtheorem*{thm*}{Theorem}

\theoremstyle{definition}

\newtheorem*{defn*}{Definition}

\makeatother

\graphicspath{{figures/}}

\usepackage{xysyntax}

\tikzstyle{none} = []
\tikzstyle{A}=[fill=none, draw=black, shape=circle, radius=20]
\tikzstyle{B}=[fill=blue, draw=blue, shape=circle, size=0.3mm]
\tikzstyle{D}=[fill=green, draw=black, shape=rectangle]
\tikzstyle{C}=[fill=red, draw=red, shape=circle, textcolor=red]
\tikzstyle{X}=[fill=cyan, draw=black, shape=circle]
\tikzstyle{new style 0}=[fill=red, draw=red, shape=circle]
\tikzstyle{new style 1}=[fill=orange, draw=orange, shape=rectangle]
\tikzstyle{grayText}=[fill=none, draw=black, shape=circle, text=black]
\tikzstyle{redText}=[fill=none, draw=none, shape=circle, text=red]
\tikzstyle{blueText}=[fill=none, draw=none, shape=circle, text=blue]
\tikzstyle{magentaText}=[fill=none, draw=none, shape=circle, text=magenta]

\tikzstyle{black}=[-, line width=0.1mm]
\tikzstyle{black2}=[-, line width=0.2mm]
\tikzstyle{magenta}=[-, draw=magenta, line width=0.1mm]
\tikzstyle{magenta2}=[-, draw=magenta, line width=0.2mm]
\tikzstyle{densely dotted}=[-, densely dotted, draw=black]
\tikzstyle{red arrows}=[<->, solid, draw=red, line width=0.5mm]
\tikzstyle{green arrow}=[<->, draw=green, line width=0.5mm]
\tikzstyle{green dashed}=[<->, dashed, draw=green, line width=0.5mm]
\tikzstyle{red solid}=[-, draw=red]
\tikzstyle{red dashed}=[-, densely dotted, draw=red]
\tikzstyle{blue solid}=[-, dashed, draw=blue, line width=0.2mm]
\tikzstyle{blue solid 2}=[-, draw=blue, line width=0.2mm]
\tikzstyle{blue dashed}=[-, densely dotted, draw=blue]
\tikzstyle{magenta arrow}=[->, draw=magenta]
\tikzstyle{dotted}=[-, dotted, draw=black]
\tikzstyle{red wavy}=[-, dashed, draw=red]
\tikzstyle{blue wavy}=[-, dashed, draw=blue]
\tikzstyle{black wavy}=[-, draw=black, line width=0.5mm]
\tikzstyle{blue arrow}=[->, draw=blue]

\begin{document}

\begin{titlepage}
\vspace*{3cm} 

\begin{center}
\vspace{1cm}
{\Large\bfseries Non-invertible Symmetries in 2D \\from Type IIB String Theory\\ 
}
\vspace{0.7cm}

{\large
Xingyang Yu$^{1,2}$\\}
\vspace{.2cm}
{ $^1$ Center for Cosmology and Particle Physics, \\New York University, New York, NY 10003, USA}\par
\vspace{.1cm}
{ $^2$ Department of Physics, Robeson Hall, Virginia Tech, \\Blacksburg, VA 24061, USA.}\par
\vspace{.1cm}

\vspace{.3cm}

\scalebox{.9}{\tt xingyangy@vt.edu}\par
\vspace{5mm}
\textbf{Abstract}
\end{center}

We propose a top-down approach to non-invertible symmetries in 2D QFTs and their 3D associated symmetry topological field theories. We focus on the gauge theory engineered on D1-branes probing a particular Calabi-Yau 4-fold singularity. We show how to derive the symmetry topological field theory, a 3D Dijkgraaf-Witten theory, from the IIB supergravity under dimensional reduction. We also identify branes behind the non-invertible topological lines by dimensionally reducing their worldvolume actions. The action of non-invertible lines on charged local operators is then realized as the Hanany-Witten transition.

\vfill 
\end{titlepage}

\tableofcontents

\newpage
\section{Introduction}
Global symmetry is one of the most important concepts in quantum field theories (QFTs). It provides powerful tools to investigate QFTs, even those strongly coupled or without Lagrangian. A modern approach to understanding global symmetries is through their associated topological symmetry operators or defects \cite{Gaiotto:2014kfa}: For a D-dimensional QFT with a $q$-form global symmetry whose symmetry group is $G$, a topological operator $U(M_{D-q-1})_{g}$ is associated with the group element $g$, and supported on the codimension-$q$ manifold $M_{D-q-1}$. An operator charged under this $q$-form symmetry is supported on $q$-dimensional manifold $N_q$, linking with the $M_{D-q-1}$. It carries a representation of the group of the group $G$, and thus transformed accordingly when acted by a topological operator $U(M_{D-q-1})_g$. The group multiplication law leads to the simple fusion rule between symmetry operators as $U(M_{D-q-1})_g\times U(M_{D-q-1})_h=U(M_{D-q-1})_{gh}$. The existence of the group element $g^{-1}$ gives rise to the invertibility of the symmetry operator: $U(M_{D-q-1})_{g} \times U(M_{D-q-1})_{g^{-1}}\equiv U(M_{D-q-1})\times U^{-1}(M_{D-q-1})=1$. Relaxing the group multiplication law and considering non-trivial fusion rules for symmetry operators $U_i$'s as $U_i(M_{D-q-1})\times U_j(M_{D-q-1})=\sum_k c_{ij}^kU_k(M_{D-q-1})$, one ends up with symmetries which are not group-like, known as \emph{non-invertible symmetries}\footnote{We refer the reader to \cite{Cordova:2022ruw, Schafer-Nameki:2023jdn, Bhardwaj:2023kri, Luo:2023ive, Shao:2023gho} for recent reviews.}.


In the context of QFTs engineered from singularities in string theory, e.g., via geometric engineering or brane probes, generalized global symmetries admit elegant top-down realizations. On the one hand, the charged defects are built by branes wrapping non-compact cycles of the internal geometry, extending from the singularity (where the QFT is engineered) to ``infinity'' \cite{DelZotto:2015isa, GarciaEtxebarria:2019caf, Albertini:2020mdx, Morrison:2020ool}. On the other hand, it is recently pointed out in \cite{Apruzzi:2022rei, GarciaEtxebarria:2022vzq, Heckman:2022muc} (see also \cite{Heckman:2022xgu, Apruzzi:2023uma, Bah:2023ymy, Dierigl:2023jdp, Cvetic:2023plv}) that generalized symmetry operators arise from wrapped branes ``at infinity''\footnote{In addition to branes, generalized symmetry operators can also arise from purely geometric configuration. See e.g., \cite{Lawrie:2023tdz} and Appendix A in \cite{Heckman:2022xgu}.}. In particular, in the case of non-invertible symmetries, the topological field theory (TFT) living on the symmetry operator, responsible for the non-trivial fusion rules, can be directly obtained from the topological sector of the brane action on its worldvolume via dimensional reduction on the wrapped cycles ``at infinity''.

Despite many top-down approaches and brane constructions for non-invertible symmetries being introduced in the literature, as far as our knowledge, they almost exclusively focus on QFTs in $D>2$ dimensions. To some extent, this is a bit surprising since non-invertible symmetries are most ubiquitous in 2D\footnote{In 2D, non-invertible symmetries have a long history. See, e.g., \cite{Verlinde:1988sn, Fuchs:2002cm, Frohlich:2004ef,  Bhardwaj:2017xup, Chang:2018iay, Thorngren:2019iar, Komargodski:2020mxz, Thorngren:2021yso} for a partial list of seminal papers.}. In this note, we fill this small gap by explicitly constructing brane origins for non-invertible symmetries in 2D QFTs with string theory realization. 

\paragraph{2D QFT on D1-branes probing singularities.}
The 2D QFTs we will focus on are gauge theories engineered on D1-branes probing the conical singularity of a Calabi-Yau 4-fold (CY$_4$). The IIB string theory background reads 
\begin{equation}
    \mathbb{R}^{1,1}\times Y,
\end{equation}
where $\mathbb{R}^{1,1}$ supports the worldvolume of a stack of $N$ D1-branes and $Y$ is a local non-compact CY$_4$. In the case when $Y$ is toric, an infinite class of 2D theories has been explicitly constructed, using an elegant T-dual IIA intersecting brane configuration known as \emph{brane brick models} \cite{Franco:2015tna, Franco:2015tya, Franco:2016nwv, Franco:2016qxh} \footnote{See \cite{Franco:2016fxm, Franco:2017cjj,  Franco:2017lpa, Franco:2018qsc, Franco:2020ijt, Franco:2021elb, Franco:2021vxq, Franco:2021ixh, Franco:2022isw} for more details.}. The resulting 2D QFTs are $U(N)^K$  quiver gauge theories\footnote{Strictly speaking, there also exist gauge theory phases whose gauge factors $U(N_i)$ can have different ranks. These are referred to as \emph{non-toric} phases \cite{Franco:2016nwv}, which can be derived by performing the $\mathcal{N}=(0,2)$ triality \cite{Gadde:2013lxa} from toric phases.}, which can be fully specified by quiver diagrams (encoding the field content and the gauge interaction) and superpotentials (encoding the matter interaction)\footnote{Brane brick models enjoy $\mathcal{N}=(0,2)$ supersymmetry. However, at the level of generalized global symmetries we discuss in this note, supersymmetry matters little.}.

\begin{figure}
    \centering
    \includegraphics[width=6cm]{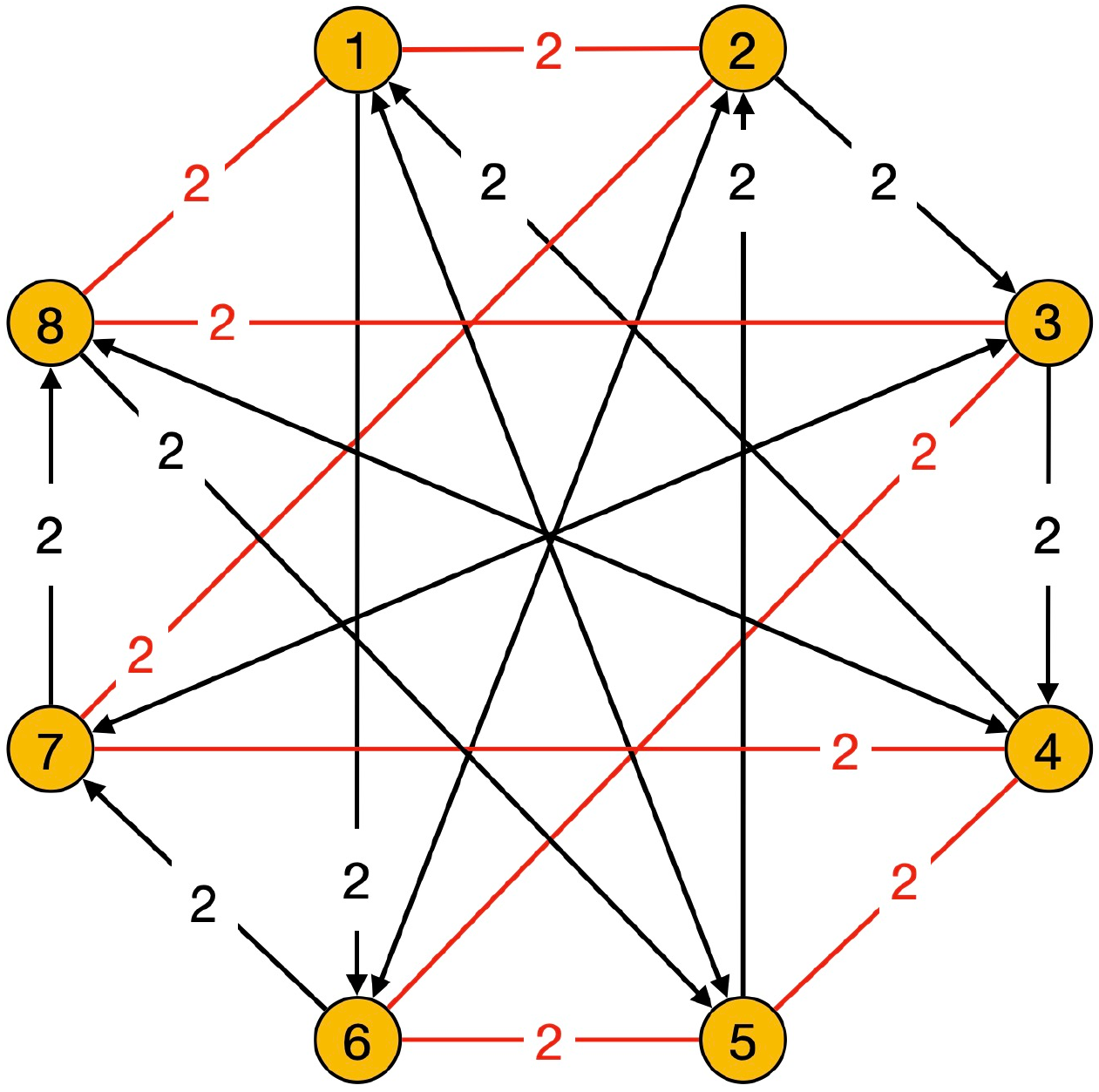}
    \caption{Quiver diagram for a 2D gauge theory phase associated with $Y^{2,0}(\mathbb{P}^1\times \mathbb{P}^1)$ probed by $N$ D1-branes \cite{Franco:2022isw}. Yellow circles denote $U(N)$ gauge groups. Oriented black lines and unoriented red lines denote bifundamental chiral and Fermi superfields, respectively.}
    \label{fig: quiver}
\end{figure}

To illustrate our idea explicitly, in this note, we focus on the 2D gauge theory associated with a specific conical CY$_4$,
\begin{equation}
  \text{Cone}(Y^{2,0}(\mathbb{P}^1\times \mathbb{P}^1)),
\end{equation}
which is the cone over a smooth 7-manifold known as $Y^{2,0}(\mathbb{P}^1\times \mathbb{P}^1)$. The 2D gauge theory is constructed in \cite{Franco:2022isw}, whose quiver diagram is shown in Figure \ref{fig: quiver}. The 7-manifold $Y^{2,0}(\mathbb{P}^1\times \mathbb{P}^1)$ falls in an infinite class of Sasaki-Einstein 7-manifolds denoted as $Y^{p,k}(\mathbb{P}^1\times \mathbb{P}^1)$, which are lens space $S^3/\mathbb{Z}_p$ bundles over $\mathbb{P}^1\times \mathbb{P}^1$ \cite{Martelli:2008rt}.  We leave the systematic treatment of non-invertible and other global symmetries for general brane brick models in the forthcoming work \cite{Franco:2023toapp}. 

\paragraph{3D symmetry TFT from string theory.}
We will use the \emph{symmetry TFT} framework to build non-invertible symmetries for our interested 2D gauge theory. Symmetry TFT is a (D+1)-dimensional TFT capturing the topological nature of generalized global symmetries in a D-dimensional QFT \cite{Reshetikhin:1991tc, Turaev:1992hq, Barrett:1993ab, Witten:1998wy, Fuchs:2002cm, Kirillov:2010nh, Kapustin:2010if, Kitaev:2011dxc, Fuchs:2012dt, Freed:2012bs, Freed:2018cec, Freed:2022qnc, Kaidi:2022cpf, Baume:2023kkf}. It has a physical boundary and a topological boundary. The local information (local operators and their correlation functions) of the interested D-dimensional QFT is realized on the physical boundary, also known as the \emph{relative QFT} \cite{Freed:2012bs}. On the other hand, gapped boundary conditions are defined on the topological boundary, which specifies the global structure of the D-dimensional QFT. 

For QFTs engineered on conical singularities of a local non-compact internal geometry $Y$ in string theory, the associated symmetry TFT can be derived from the topological sector of the dimensional reduction for the 10D (11D for M-theory) supergravity on the asymptotic boundary $\partial Y$ \cite{Apruzzi:2021nmk} (see also \cite{vanBeest:2022fss, Heckman:2022xgu, Etheredge:2023ler, Lawrie:2023tdz, GarciaHosseini, Apruzzi:2023uma, Baume:2023kkf}). Various string theory fluxes under dimensional reduction give rise to gauge fields in the symmetry TFT. For our interested case in this note, namely, D1-branes probing Cone$(Y^{(2,0)}(\mathbb{P}^1\times \mathbb{P}^1))$, the dimensional reduction to obtain a 3D symmetry TFT is  performed in the IIB string theory background 
\begin{equation}
    M_3\times Y^{(2,0)}(\mathbb{P}^1\times \mathbb{P}^1),
\end{equation}
where $M_3\cong M_2\times \mathbb{R}_{r \geq 0}$ is the 3D manifold for the symmetry TFT bulk. The physical boundary corresponds to $r=0$ where D1-branes are localized, while the topological boundary arises at $r=\infty$ where boundary conditions of various IIB fluxes are picked. The detailed computation will be discussed in Section 2, where we show the resulting 3D symmetry TFT is a twisted $\mathbb{Z}_2\times \mathbb{Z}_2\times \mathbb{Z}_2$ 3D Dijkgraaf-Witten theory
\begin{equation}\label{eq:intro 3D DW theory}
\boxed{
 S_3=\frac{2\pi}{2}\int_{M_3} a_1 \delta \hat{a}_1+b_1\delta \hat{b}_1+c_1\delta \hat{c}_1+a_1b_1c_1,
}
\end{equation}

\paragraph{Non-invertible symmetry operators from branes.}
Recall that picking a topological boundary condition for the symmetry TFT corresponds to fixing a global structure of its associated D-dimensional QFT. This procedure is called picking a \emph{polarization}, and the resulting QFT with a well-defined global structure is referred to as an \emph{absolute QFT} (see, e.g.,  \cite{Belov:2006jd, Freed:2012bs, Gukov:2020btk, Lawrie:2023tdz}). From this bulk perspective, gauging a symmetry in a QFT to get another QFT is translated in changing from one polarization to another.

Global symmetries of the resulting absolute QFT can be obtained by investigating the behavior of bulk operators under the topological boundary condition. Operators trivialized when touching the gapped boundary (due to the possible Dirichlet condition), giving rise to charged defects. In contrast, those not trivialized are still topological operators, generating global symmetries for the absolute QFT.

Based on this general idea, one can start with the symmetry TFT (\ref{eq:intro 3D DW theory}) and write down gauge-invariant line operators by purely field-theoretic consideration, much as in \cite{deWildPropitius:1995cf, Kaidi:2023maf}. However, since the Dijkgraaf--Witten theory (\ref{eq:intro 3D DW theory}) is derived from string theory, one naturally asks whether there is a direct top-down approach to topological line operators. The answer is indeed yes. As we will discuss in Section 2, all line operators, whether invertible or not, have their corresponding brane origin. Line operators are obtained exactly from the brane worldvolume action via dimensional reduction on various cycles wrapped by branes.

Having obtained line operators in the 3D bulk from branes, building non-invertible symmetries in 2D gauge theory associated with $Y^{(2,0)}(\mathbb{P}^1\times \mathbb{P}^1)$ then translates in writing down polarizations under which the non-invertible bulk lines are still non-invertible when touching the gapped boundary. We will show in Section 3 that these polarizations indeed exist, and the resulting non-invertible symmetry is the well-known $\mathbb{Z}_2\times \mathbb{Z}_2$ Tambara-Yamagami fusion category \cite{TAMBARA1998692}! For example, the polarization corresponding to the boundary condition 
\begin{equation}\label{eq: intro non-standard tbc}
   a_1, \hat{b}_1, \hat{c}_1~\text{Dirichlet}; ~\hat{a}_1,b_1, c_1~\text{Neumann},
\end{equation}
has the following non-invertible fusion rules
\begin{equation}
\boxed{
\begin{split}
	&\mathcal{N}_\text{D3}\times \mathcal{N}_\text{D3}=1+\eta_{\text{F1}}+\eta_{\text{D1}}+\eta_{\text{F1}}\eta_{\text{D1}},\\
	&\eta_{\text{F1}}\times \eta_{\text{F1}}=\eta_{\text{D1}}\times \eta_{\text{D1}}=1,\\
	&\eta_{\text{F1}}\times \mathcal{N}_\text{D3}= \eta_{\text{D1}}\times \mathcal{N}_\text{D3}=\mathcal{N}_\text{D3},
\end{split}
}
\end{equation}
where $\mathcal{N}_{\text{D3}}$ is the non-invertible line from D3-brane, while $\eta_{\text{F1}}, \eta_{\text{D1}}$ are invertible $\mathbb{Z}_2$ lines from F1- and D1-strings respectively.

In addition to polarizations enjoying non-invertible symmetries, we also find polarizations where all topological line operators become invertible symmetry lines. That is to say, the non-invertible symmetries we construct in this note are \emph{non-intrinsic} \cite{Kaidi:2022uux, Bashmakov:2022uek, Sun:2023xxv}.

\section{3D Dijkgraaf--Witten Theory and its Line Operators from IIB}
In this section, we present how to obtain the 3D symmetry TFT and its line operators for 2D gauge theory associated with Cone$(Y^{2,0}(\mathbb{P}^1\times \mathbb{P}^1))$ from IIB string theory via dimensional reduction. In particular, we find the following top-down approach to the field theory content
\begin{equation}
\begin{split}
     \text{IIB supergravity} &\longrightarrow \text{3D Dijkgraaf--Witten theory},\\
    \text{branes worldvolume actions} &\longrightarrow \text{Line operators in the 3D bulk}
\end{split}
\end{equation}

\subsection{3D Dijkgraaf--Witten theory from the IIB supergravity}

To derive the 3D symmetry TFT, we focus on the topological sector of the reduction for IIB string theory on the asymptotic boundary of the Calabi-Yau 4-fold, which in this case is just the base manifold $L_7\equiv Y^{2,0}(\mathbb{P}^1\times \mathbb{P}^1)$ at infinity. In particular, we treat the various IIB supergravity fluxes as elements in differential cohomology uplifts of (see, e.g.,\cite{Belov:2006jd, Apruzzi:2021nmk} )
\begin{equation}\label{eq:coho class of L7}
    H^*(L_7,\mathbb{Z})=\left\{ \mathbb{Z},0,\mathbb{Z}^2\oplus \mathbb{Z}_2, 0, \mathbb{Z}\oplus \mathbb{Z}_2\oplus \mathbb{Z}_2, \mathbb{Z}^2, \mathbb{Z}_2, \mathbb{Z} \right\}.
\end{equation} 
The above cohomology classes for $L_7=Y^{2,0}(\mathbb{P}^1\times \mathbb{P}^1)$ can be found in \cite{Martelli:2008rt}.

The relevant topological action inherited from the IIB string theory, roughly speaking, consists of two parts. The quadratic part comes from the kinetic terms for IIB fluxes, and the cubic part comes from the 10D Chern-Simons coupling $-\int C_4 \wedge dB_2 \wedge dC_2$.\footnote{The topological action of IIB string theory for symmetry TFT computation has been investigated in, e.g., \cite{Heckman:2017uxe, GarciaEtxebarria:2019caf, Apruzzi:2021nmk,  Apruzzi:2022dlm, Lawrie:2023tdz, Baume:2023kkf}. We also refer the reader to the recent work \cite{Apruzzi:2023uma} for a more systematic discussion.} Consider an IIB string theory background without 7-branes. The topological action that we start with reads 
\begin{equation}\label{eq:11D quadra action}
    \frac{S_{11}}{2\pi}=\int_{N_4\times L_7}\frac{1}{2}\breve{F}_6\star \breve{F}_6-\breve{F}_6\star \breve{H_3} \star \breve{G}_3
\end{equation}
which lives in 11D spacetime $N_4\times L_7$. The 4-manifold $N_4$ satisfies $\partial N_4=M_3$, an auxiliary bulk manifold whose boundary is the 3-manifold where the symmetry TFT lives. Note that all terms are 12-dimensional since we have uplifted IIB fluxes as differential cohomology elements. $\breve{F}_6$ is the differential cohomology element whose connection part is the IIB self-dual D3-brane 5-form flux $F_5$. $\breve{H_3}$ and $\breve{G}_3$ are differential cohomology uplift for F1- and D1-string flux $dB_2$ and $dC_2$.\footnote{The $\star$ symbol defines a bilinear product operation on Cheeger--Simons characters  $\breve{H}^{k_1}(M_d)\times \breve{H}^{k_2}(M_d)=\breve{H}^{k_1+k_2}(M_d)$ \cite{brylinski2007loop, Cheeger1985DifferentialCA}. In particular, when $k_1+k_2=d+1$, the integral describes a perfect pairing $\breve{H}^{k_1}(M_d)\times \breve{H}^{d+1-k_1}(M_d)\rightarrow \mathbb{R}/\mathbb{Z}$. We refer the reader to \cite{baer2013differential} for a nice review of differential cohomology.}

According to (\ref{eq:coho class of L7}), we expand differential cohomology elements as
\begin{equation}\label{eq:expansion fluxes for quadra}
\begin{split}
    \breve{F}_{6}=&\breve{f}_6\star \breve{1}+\sum_{\alpha=1}^2\breve{F}_4^{(\alpha)}\star \breve{u}_{2(\alpha)}+\breve{F}_2\star \breve{u}_4+\sum_{\alpha=1}^2\breve{F}_{1(\alpha)}\star \breve{u}_{5}^{(\alpha)}
    +\breve{A}_4\star \breve{t}_2+\sum_{i=1}^2\breve{A}_2^{(i)}\star \breve{t}_{4(i)},\\
    \breve{G}_{3}=&N\breve{\text{vol}}_{M_3}\star \breve{1}+\sum_{\alpha=1}^2\breve{G}_1^{(\alpha)}\star \breve{u}_{2(\alpha)}
    +\breve{C}_1\star \breve{t}_2,\\
    \breve{H}_{3}=&\breve{h}_3\star \breve{1}+\sum_{\alpha=1}^2\breve{H}_1^{(\alpha)}\star \breve{u}_{2(\alpha)}
    +\breve{B}_1\star \breve{t}_2,\\
\end{split}
\end{equation}
where the generators for various cohomology classes are denoted as  
\begin{equation}
\begin{split}
    \breve{1} &\leftrightarrow H^0(L_7,\mathbb{Z})=\mathbb{Z},\\
    \breve{u}_{2(\alpha)},\alpha=1,2 &\leftrightarrow \text{non-torsional}~H^2(L_7,\mathbb{Z})=\mathbb{Z}^2,\\
    \breve{u}_{4} &\leftrightarrow \text{non-torsional}~H^4(L_7,\mathbb{Z})=\mathbb{Z},\\
    \breve{u}_{5}^{(\alpha)},\alpha=1,2 &\leftrightarrow H^5(L_7,\mathbb{Z})=\mathbb{Z}^2,\\
    \breve{\text{vol}} &\leftrightarrow H^{7}(L_7,\mathbb{Z})=\mathbb{Z},\\
    \breve{t}_2 &\leftrightarrow \text{torsional}~H^2(L_7,\mathbb{Z})=\mathbb{Z}_2,\\
    \breve{t}_{4(i)}, i=1,2 &\leftrightarrow \text{torsional}~H^4(L_7,\mathbb{Z})=\mathbb{Z}_2\oplus \mathbb{Z}_2,\\
    \breve{t}_6 &\leftrightarrow H^6(L_7,\mathbb{Z})=\mathbb{Z}_2.
\end{split}
\end{equation}
Fields from torsional parts give rise to finite symmetries while those from non-torsional parts correspond to continuous symmetries. In this work, we only focus on finite symmetries and their descendent non-invertible symmetries\footnote{For brane interpretation of continuous symmetry operators, we refer the reader to \cite{Cvetic:2023plv}.}, so we only turn on the fields as coefficients of the torsional generators $t_p$, where $p=2,4,6$.

Substituting the torsional part of (\ref{eq:expansion fluxes for quadra}) into the 11D topological action (\ref{eq:11D quadra action}), we derive the 3D symmetry TFT for finite symmetries,
\begin{equation}\label{eq: 3D TFT part 1}
\begin{split}
    \frac{S_{3}}{2\pi}&=\int_{N_4}\sum_{i,j=1}^2\Lambda_{ij}\breve{A}_2^{(i)}\star \breve{A}_2^{(j)}- \sum_{i}\Delta_i \breve{A}_2^{(i)}\star \breve{B}_1 \star \breve{C}_1\\
    &=\int_{M_3}\sum_{i,j}\Lambda_{ij}a_1^{(i)}\delta a_1^{(j)} - \sum_{i=1}^2\Delta_i a_1^{(i)}b_1c_1
\end{split}
\end{equation}
where we use respective lower-case letters to express fields in terms of the ordinary cohomology elements and omit the ``$\cup$'' product symbol for simplicity. The coefficients in the action are given by the linking numbers within the 7-manifold $L_7$,
\begin{equation}\label{eq: linking numbers}
\begin{split}
     \Lambda_{ij}&\equiv \frac{1}{2}\int_{L_7}\breve{t}_{4(i)}\star \breve{t}_{4(j)}~\text{mod 1},\\
 \Delta_i&\equiv \int_{L_7}\breve{t}_{4(i)}\star \breve{t}_2 \star \breve{t}_2~\text{mod 1},
\end{split}
\end{equation}
whose derivation requires expressing $p$-dimensional torsional generators $\breve{t}_p$ in terms of  various compact $(8-p)$-cycles in the toric Calabi-Yau 4-fold \cite{vanBeest:2022fss}. The linking number computation then translates into reading quadruple intersection numbers between codimension-2 divisors in the toric varieties\footnote{See Chapter 7 in \cite{hori2003mirror} for how to compute intersection numbers in toric varieties.}. 

It is easy to see the action (\ref{eq: 3D TFT part 1}) is not complete. Notice that the quadratic term comes from the non-commutativity for the boundary profile of the self-dual 5-form; thus, other non-commutative fluxes should also be captured in the resulting 3D TFT \cite{GarciaEtxebarria:2019caf}. This leads to adding quadratic terms for $b_1$ and $c_1$ from the non-commutativity F1-NS5 and D1-D5 pairs when wrapping torsional cycles linking to each other \footnote{We thank Inaki Garcia Etxebarria for valuable discussions on this point}. The resulting TFT reads 
\begin{equation}
\begin{split}
    \frac{S_{3}}{2\pi}=\int_{M_3}\sum_{i,j}\Lambda_{ij}a_1^{(i)}\delta a_1^{(j)}+\Omega \left( -c_1\delta \hat{c_1}+b_1\delta \hat{b}_1 \right) - \sum_{i=1}^2\Delta_i a_1^{(i)}b_1c_1
\end{split}
\end{equation}
where $\hat{c}_1$ and $\hat{b}_1$ are from IIB fluxes $\hat{G}_7$ and $\hat{H}_7$ (10D Hodge-dual of $G_3$ and $H_7$) via reduction on torsional 5-cycles $\gamma_5$ associated to the generator $\breve{t}_6$\footnote{We would like to stress that differential cohomology and flux non-commutativity is not the only way to read the quadratic terms in the symmetry TFT. In fact, it is also possible to derive these terms directly from the supergravity kinetic terms. See \cite{Camara:2011jg, GarciaHosseini}, and Appendix B in \cite{Baume:2023kkf} for more details.}. $\Omega$ denotes the  linking number between the $\breve{t}_2$ and $\breve{t}_6$ generators:
\begin{equation}\label{eq: linking number 2}
     \Omega \equiv \int_{L_7} \breve{t}_2 \star \breve{t}_6~\text{mod 1}.
\end{equation}

Computing the linking number (\ref{eq: linking numbers}, \ref{eq: linking number 2}) and redefining the notation as 
\begin{equation}
    a_1^{(1)}\rightarrow a_1,~a_1^{(2)}\rightarrow \hat{a}_1,
\end{equation}
we end up with an elegant result
\begin{equation}\label{eq:3D DW theory}
\boxed{
 S_3=\frac{2\pi}{2}\int_{M_3} a_1 \delta \hat{a}_1+b_1\delta \hat{b}_1+c_1\delta \hat{c}_1+a_1b_1c_1,
}
\end{equation}
which is just a 3D $\mathbb{Z}_2\times \mathbb{Z}_2\times \mathbb{Z}_2$ Dijkgraaf-Witten theory with a simple twist $a_1b_1c_1$.

Each TFT field serves as the background gauge field for a factor within the defect group \cite{DelZotto:2015isa, Lawrie:2023tdz}, which now can be straightforwardly read as 
\begin{equation}\label{eq: defect group}
	\mathbb{D}=\left( \mathbb{Z}^{a}_2\times \mathbb{Z}^{\hat{a}}_2 \right)\oplus \left( \mathbb{Z}^{b}_2\times \mathbb{Z}^{\hat{b}}_2 \right)\oplus \left( \mathbb{Z}^{c}_2\times \mathbb{Z}^{\hat{c}}_2 \right).
\end{equation}
We use ``$\times$'' to denote the group factors with non-trivial Dirac pairing between their defects, or equivalently, those background gauge fields canonical conjugate to each other under the TFT quantization. ``$\oplus$'', on the contrary, means group factors without any non-commutativity between their fluxes.

\subsection{Line operators from brane worldvolume actions}
Having derived a 3D Dijkgraaf-Witten theory as the symmetry TFT for finite symmetries in the 2D QFT, the next natural question is: What is the spectrum of line operators in this 3D TFT, and how do these operators translate in (non-invertible) topological defect lines in the 2D QFT? Field-theoretically, this question has been intensively investigated in, e.g., \cite{Kaidi:2023maf}. In this section, we will provide a top-down treatment where topological defect lines, no matter whether invertible or not, enjoy elegant origins as branes in the IIB string theory.

The first step is to determine the candidate of branes that are responsible for line operators in the Dijkgraaf-Witten theory (\ref{eq:3D DW theory}). Recall finite gauge fields in (\ref{eq:3D DW theory}) are reduced from various IIB fluxes, each of which couples to a certain type of branes. For example, $\hat{c}_1$ is the expansion field from the $\breve{t}_6$ reduction of $\breve{\hat{G}}_7$ (as a differential cohomology element) on $\text{Tor}H^6(L_7,\mathbb{Z})$, which corresponds to the IIB flux $\hat{G}_7$ (electrically) coupled to D5-branes. More precisely, according to the universal coefficient theorem $\text{Tor}H_n=\text{Tor}H^{n+1}$, we have the correspondence between torsional cohomology generators and torsional cycles 
\begin{equation}
	\gamma_{n}^{(i)}\leftrightarrow t_{n+1(i)}.
\end{equation}
This translates in the $\hat{c}$ case as  
\begin{equation}
	\hat{c} ~~\leftrightarrow~ \text{D5s on }\gamma_{5},
\end{equation} 
where $\gamma_5\in \text{Tor}H_5(L_7,\mathbb{Z})$ is the torsional 5-cycle dual to the $\breve{t}_6$ generator. Similarly, one can derive brane patterns associated with each TFT field. The dimensional reduction of the D5-brane topological coupling then gives the corresponding naive magnetic line operator in the 3D TFT:
\begin{equation}\label{eq:D5 WZ leading term}
	\exp{ \left(2\pi i \int_{M_6} C_6 \right)}\rightarrow \exp\left( 2\pi i\int_{M_1\times \gamma_5}\breve{\hat{G}}_7 \right)=\exp\left( \pi i\int_{M_1}\hat{c}_1 \right).
\end{equation}

However, this invertible line is not the full construction of the magnetic operator dependent on $\hat{c}$, because it is not gauge-invariant within our interested Dijkgraaf-Witten theory (see also e.g., \cite{Kaidi:2023maf}). Note that $C_6$ does not carry the full topological information of the D5-brane but only the leading term of the Wess-Zumino part of the D5-brane action. In order to encode the full topological effect of the D5-brane on the 3D Dijkgraaf-Witten theory (or equivalently, on the 2D QFT on the physical boundary of the 3D bulk), we consider the following action,
\begin{equation}
	S^{\text{top}}_{\text{D5}}=\int\mathcal{D}\hat{f}_4\mathcal{D}f_2 \exp \left( 2\pi i\int_{N_2\times \gamma_5}\hat{f}_4d f_2+\hat{G}_7-F_5(B_2-f_2)-\frac{1}{2}G_3(B_2-f_2)^2 \right).
\end{equation}
This is a topological action on an auxiliary 7D bulk $N_2\times \gamma_5$, where $N_2$ satisfies $\partial N_2=M_1$, i.e. an auxiliary 2-manifold whose boundary is the topological line supporting the operator in the resulting 3D TFT. In this topological action, $f_2$ is the field strength of the dynamical gauge field from the F1 open string fluctuation, and $\hat{f}_4$ is its hodge dual on the D5-brane worldvolume. The first term thus carries the relevant information from the Dirac-Born-Infeld part of the D-brane action \cite{Apruzzi:2023uma}. The other three terms come from the Wess-Zumino part of the brane action \cite{Douglas:1995bn}, where the leading term $\hat{G}_7$ is the origin for the naive magnetic operator which we discussed in (\ref{eq:D5 WZ leading term}). $F_5, G_3$ are fluxes for the induced lower dimensional D3- and D1-brane charges, respectively, while $B_2$ is the regular notation for the NS-NS field electrically coupled to F1-strings. Note that the path integral is only performed over $f_2$ and $\hat{f}_4$, which are dynamical degrees of freedon on the D5-brane worldvolume.

In order to perform the dimensional reduction on the torsional cycle $\gamma_5$, as what we did in computing the symmetry TFT, we promote the topological action in terms of the differential cohomology elements\footnote{The $(\breve{H}_3-\breve{f}_3)^2$ here means an order-5 differential cohomology element from the star product between the differential cohomology element $\breve{H}_3-\breve{f}_3$ and its connection part.}
\begin{equation}
	S_{\text{D5}}^{\text{top}}\rightarrow \int \mathcal{D}\breve{\hat{f}}_5\mathcal{D}\breve{f}_3 \exp \left( 2\pi i\int_{N_2\times \gamma_5}\breve{\hat{f}}_5 \star \breve{f_3}+\breve{\hat{G}}_8-\breve{F}_5\star (\breve{H}_3-\breve{f}_3)-\frac{1}{2}\breve{G}_3\star (\breve{H}_3-\breve{f}_3)^2 \right).
\end{equation}
The expansions of $ \breve{F}_5$ and $\breve{H}_3$ are already given in (\ref{eq:expansion fluxes for quadra})\footnote{Note that in (\ref{eq:expansion fluxes for quadra}), the $\breve{F}_6$ is the differential cohomology element via $F_5$ is its connection part, but here $\breve{F}_5$ is the differential cohomology element itself, regarded as gauge-invariant field strength.}, while the expansion for $\breve{\hat{G}}_8, \breve{\hat{f}}_5$ and $\breve{f}_3$ can be defined as
\begin{equation}
\begin{split}
\breve{\hat{f}}_5&=\hat{\phi}_1\star \breve{t}_{4(1)}+\breve{\phi}_{1}^\prime \star \breve{t}_{4(2)}+\cdots,\\
\breve{f}_3&=\breve{\phi}_1\star \breve{t}_2+ \cdots,
\end{split}
\end{equation}
where we only write down terms relevant to the reduction on the torsional cycle $\gamma_5$. Again, using the linking number between various cohomology generators, the resulting 1D TFT reads
\begin{equation}\label{eq: D5 as magnetic line}
	S_{\text{D5}}^{\gamma_5}\propto \int \mathcal{D}\hat{\phi}_0 \mathcal{D}\phi_0 \exp \left( \pi i \int_{M_1} \hat{c}_1 \right)  \exp \left( \pi i \int_{M_1} \hat{\phi}_0\delta \phi_0+\phi_0 a_1-\phi_0 b_0 c_1+\frac{1}{2}\phi_0^2c_1 \right),
\end{equation}
where $db_0=b_1$, and we have omitted all other terms decoupled from the dynamical $\hat{\phi}_0$ and $\phi_0$. Taking variation of $\phi_0$, we get the condition
\begin{equation}
	\delta \hat{\phi}_0=a_1+\phi_0c_1-b_0c_1,
\end{equation}
substituting which back to (\ref{eq: D5 as magnetic line}), we integral over $c_1$ and end up with the topological line operator
\begin{equation}
	S_{\text{D5}}^{\gamma_5}\propto \mathcal{N}_{\text{D5}}(M_1)\equiv \int\mathcal{D}\hat{\phi}_0 \mathcal{D}\phi_0 \exp \left( \pi i \int_{M_1} \hat{c}_1 \right)  \exp \left( \pi i \int_{M_1} \hat{\phi}_0\delta \phi_0+\phi_0 a_1-\hat{\phi}_0 b_1 \right).
\end{equation}
This is a non-invertible gauge-invariant magnetic line operator for the 3D Dijkgraaf-Witten theory, matching the result in \cite{Kaidi:2023maf}.

The fusion rule for this line operator is 
\begin{equation}
\begin{split}
	&\mathcal{N}_{\text{D5}}\times \mathcal{N_\text{D5}}=\left[ 1+ \exp\left( \pi i\int_{M_1}a_1 \right)\right] \left[ 1 +\exp\left( \pi i\int_{M_1}b_1\right) \right].\\
\end{split}
\end{equation}
Further note that $e^{\pi i \int_{M_1} a_1}\equiv \eta_{a}$ and $e^{\pi i \int_{M_1} b_1}\equiv\eta_{b}$ are $\mathbb{Z}_2$ topological lines, where we use $\eta$ to denote invertible lines with subindices showing the TFT field dependence. The right-hand side of the above equation is the condensation of $\mathbb{Z}_2\times \mathbb{Z}_2$ topological lines on the $M_1$, which can be regarded as a result of higher-gauging \cite{Roumpedakis:2022aik}. Now we can write down the full fusion rule involving non-invertible line $\mathcal{N}_{D5}$, invertible $\mathbb{Z}_2$ lines $\eta_{a}$ and $\eta_{b}$ as 
\begin{equation}\label{eq:fusion of D5 operator}
\begin{split}
	&\mathcal{N}_\text{D5}\times \mathcal{N}_\text{D5}=1+\eta_{a}+\eta_{b}+\eta_{a}\eta_{b},\\
	&\eta_{a}\times \eta_{a}=\eta_{b}\times \eta_{b}=1,\\
	&\eta_{a}\times \mathcal{N}_\text{D5}= \eta_{b}\times \mathcal{N}_\text{D5}=\mathcal{N}_\text{D5},
\end{split}
\end{equation}
which is exactly the $\mathbb{Z}_2\times \mathbb{Z}_2$ Tambara-Yamagami category \cite{TAMBARA1998692}.

 Similarly, we can derive other topological line operators respectively dependent on $a_1, \hat{a}_1, b_1,\hat{b}_1$ and $c_1$ as what we did for $\hat{c}$ and its corresponding D5-brane action. We leave the computation to the interested reader as an exercise and conclude the results in Table \ref{tab: branes behind lines}.

\begin{table}[H]
    \centering
    \begin{tabular}{|c|c|}
\hline
 Line Operators in 3D TFT & Branes Configuration \\
     \hline
      $\eta_{\text{D3}}=e^{ \pi i \int_{M_1}a_1 }$ & D3-brane on $\gamma_3^{(1)}$ \\
      \hline
      $\eta_{\text{F1}}=e^{ \pi i \int_{M_1}b_1 }$ & F1-string on $\gamma_1$ \\
      \hline
      $\eta_{\text{D1}}=e^{ \pi i \int_{M_1}c_1 }$ & D1-string on $\gamma_1$ \\
      \hline
      $\mathcal{N}_{\text{D3}}= \int\mathcal{D}\hat{\phi}_0 \mathcal{D}\phi_0 e^{ \pi i \int_{M_1} \hat{a}_1 }  e^{ \pi i \int_{M_1} \hat{\phi}_0\delta \phi_0+\phi_0 b_1-\hat{\phi}_0 c_1 }$ &  D3-brane wrapping $\gamma_{3}^{(2)}$ \\
      \hline
 $\mathcal{N}_{\text{NS5}}= \int\mathcal{D}\hat{\phi}_0 \mathcal{D}\phi_0 e^{ \pi i \int_{M_1} \hat{b}_1 }  e^{ \pi i \int_{M_1} \hat{\phi}_0\delta \phi_0+\phi_0 c_1-\hat{\phi}_0 a_1 }$ &  NS5-brane wrapping $\gamma_5$\\
      \hline
      $\mathcal{N}_{\text{D5}}= \int\mathcal{D}\hat{\phi}_0 \mathcal{D}\phi_0 e^{ \pi i \int_{M_1} \hat{c}_1 }  e^{ \pi i \int_{M_1} \hat{\phi}_0\delta \phi_0+\phi_0 a_1-\hat{\phi}_0 b_1 }$ &  D5-brane wrapping $\gamma_5$\\
      \hline
\end{tabular}
\caption{Line operators in 3D Dijkgraaf-Witten theory (\ref{eq:3D DW theory}) and their brane origins. The first three brane configurations give rise to invertible electric lines, while the last three wrapped branes correspond to non-invertible magnetic lines.}
\label{tab: branes behind lines}
\end{table}

It is easy to see the electric $\mathbb{Z}_2$ lines in (\ref{eq:fusion of D5 operator}) are identified with the brane origin $\eta_{a}=\eta_{\text{D3}}, \eta_{b}=\eta_{\text{F1}}$. Furthermore, the non-invertible lines from D3- and NS5-branes also obey the $\mathbb{Z}_2\times \mathbb{Z}_2$ Tambara-Yamagami fusion category respectively:
\begin{equation}\label{eq:fusion of D3 operator}
\begin{split}
	&\mathcal{N}_\text{D3}\times \mathcal{N}_\text{D3}=1+\eta_{\text{F1}}+\eta_{\text{D1}}+\eta_{\text{F1}}\eta_{\text{D1}},\\
	&\eta_{\text{F1}}\times \eta_{\text{F1}}=\eta_{\text{D1}}\times \eta_{\text{D1}}=1,\\
	&\eta_{\text{F1}}\times \mathcal{N}_\text{D3}= \eta_{\text{D1}}\times \mathcal{N}_\text{D3}=\mathcal{N}_\text{D3},
\end{split}
\end{equation}
and 
\begin{equation}\label{eq:fusion of NS5 operator}
\begin{split}
	&\mathcal{N}_\text{NS5}\times \mathcal{N}_\text{NS5}=1+\eta_{\text{F1}}+\eta_{\text{D3}}+\eta_{\text{F1}}\eta_{\text{D3}},\\
	&\eta_{\text{F1}}\times \eta_{\text{F1}}=\eta_{\text{D3}}\times \eta_{\text{D3}}=1,\\
	&\eta_{\text{F1}}\times \mathcal{N}_\text{NS5}= \eta_{\text{D3}}\times \mathcal{N}_\text{NS5}=\mathcal{N}_\text{NS5},
\end{split}
\end{equation}

\section{Branes Behind Polarizations and Non-invertible Symmetries in 2D}

The 3D symmetry TFT bulk itself does not fully specify the global symmetry structure of the 2D QFT. At this stage, the 2D QFT associated with the conical singularity probed by D1-branes is a relative QFT \cite{Freed:2012bs, Gukov:2020btk, Lawrie:2023tdz}. It does not have a well-defined scalar-valued partition function but carries a partition vector. The corresponding space for the partition vector is regarded as the Hilbert space $\mathcal{H}$ from the 3D TFT quantization (see, e.g., \cite{Freed:2012bs, Witten:1998wy, Fuchs:2002cm}). Therefore, in this sense, the 2D QFT is ``relative" to the 3D bulk theory. 

In order to get rid of the ``relativeness'' upon the 3D bulk and thus obtain a well-defined QFT with a scalar-valued partition function, we need to pick a polarization for the system. From the 3D TFT perspective, this translates in introducing a purely gapped boundary, on which we impose a topological boundary condition. Such a boundary condition can be equivalently presented as a Lagrangian subgroup $L\subset \mathbb{D}$ of the defect  group $\mathbb{D}$.\footnote{Mathematically, the partition vector space of a relative QFT is captured by the Heisenberg group $\underline{H}^1(M_{2},\mathbb{D})$ with coefficients in the defect group $\mathbb{D}$. Picking a polarization corresponds to picking a maximally isotropic subspace of the Heisenberg group. We refer the interested reader to \cite{Lawrie:2023tdz} for a detailed discussion.} With respect to the partition vector space under the 3D TFT quantization, the relative QFT
and the gapped boundary condition can be expressed as two boundary states

Colliding the gapped boundary with the relative QFT boundary, one obtains a genuine 2D system, known as an absolute QFT, that enjoys a scalar-valued partition function. This process can be nicely expressed in terms of the inner product between boundary states $|\mathcal{R}\rangle$ and $|\mathcal{P};B\rangle $ in the partition vector space $\mathcal{H}$:
\begin{equation}
	Z_{\mathcal{P}}[B]=\langle \mathcal{R}|\mathcal{P};B\rangle.
\end{equation}
In this expression, $\langle \mathcal{R}|$ denotes the relative QFT (dual) partition vector, $|\mathcal{P};B\rangle$ denotes the boundary state for polarization $\mathcal{P}$ with the flux profile $B$, and $Z_\mathcal{P}[B]$ gives rise to the well-defined partition function with the presence of the background $B$\footnote{We remark that picking polarizations is not always possible for a generic relative QFT. Well-known examples of this type include many 2D chiral CFTs and 6D SCFTs. See, e.g., \cite{Freed:2012bs, Lawrie:2023tdz} for more details.}.

\subsection{A ``Standard'' polarization with only invertible symmetries}
Come back to the 3D Dijkgraaf-Witten theory (\ref{eq:3D DW theory}) and its relative 2D QFT associated with $Y^{2,0}(\mathbb{P}^1\times \mathbb{P}^1)$. The simplest boundary condition one can consider is 
\begin{equation}\label{eq: standard tbc}
   a_1, b_1, c_1~\text{Dirichlet}; ~\hat{a}_1,\hat{b}_1, \hat{c}_1~\text{Neumann}.
\end{equation}
This corresponds to the polarization which picks the Lagrangian subgroup $L$ of the defect group (\ref{eq: defect group})
\begin{equation}
	L=\mathbb{Z}^{\hat{a}}_2\times \mathbb{Z}^{\hat{b}}_2 \times \mathbb{Z}^{\hat{c}}_2.
\end{equation}
Therefore, the resulting absolute 2D theory has a $(\mathbb{Z}_2)^3$ global symmetry 
\begin{equation}\label{eq: standard global symmetry}
	G=\mathbb{Z}_2^{a}\times \mathbb{Z}^b_2 \times \mathbb{Z}^c_2.
\end{equation}
Based on their behavior under the gapped boundary condition (\ref{eq: standard tbc}), line operators in the 3D TFT shown in Table \ref{tab: branes behind lines} induce to various local charged operators and topological defect lines in the 2D absolute theory. 

For instance, due to the Dirichlet condition of $a_1$, $\eta_{\text{D3}}$ will terminate on the gapped boundary, i.e., it does not continue to fluctuate along the boundary and thus becomes a local operator after shrinking the 3D TFT bulk. On the contrary, $\mathcal{N}_{\text{D3}}$ is not fully trivialized on the gapped boundary due to the Neumann condition of $\hat{a}_1$. Its line manifold continues along the gapped boundary and thus gives rise to a topological defect line. However, it loses its non-invertible property during this process. To see this, notice that $b_1$ and $c_1$ are trivialized on the gapped boundary, leading to 
\begin{equation}
	\mathcal{N}_{\text{D3}}\rightarrow e^{\pi i\int_{M_1}\hat{a}_1}\int\mathcal{D}\hat{\phi}_0 \mathcal{D}\phi_0  e^{ \pi i \int_{M_1} \hat{\phi}_0\delta \phi_0 }\propto e^{\pi i\int_{M_1}\hat{a}_1},
\end{equation}
where the path integral over $\phi_0$ and $\hat{\phi}_0$ is now totally decoupled as a overall factor. Therefore, $\mathcal{N}_{\text{D3}}$, under this polarization/absolute QFT, becomes an invertible $\mathbb{Z}_2$ line. 

What is the brane configuration behind all this?  Recall that the brane origins of $\eta_{\text{D3}}$ and $\mathcal{N}_{\text{D3}}$ are D3-branes wrapping $\gamma_3^{(1)}$ and $\gamma_{3}^{(2)}$, respectively. The non-trivial linking between these two torsional 3-cycles is responsible for the canonical conjugation between $a_1$ and $\hat{a}_1$ in the 3D Dijkgraaf-Witten theory, thus tells us the local operator reduced from $\eta_{\text{D3}}$ is charged under the $\mathbb{Z}_2^{a}\subset L^\vee$ symmetry generated by the invertible line reduced from $\mathcal{N}_{\text{D3}}$. In the 10D IIB string theory picture, the gapped boundary for the 3D TFT is translated into the topological boundary conditions on the asymptotic boundary $Y^{2,0}(\mathbb{P}^1\times \mathbb{P}^1)$ ``at infinity''. Therefore, we have the following correspondence between the brane pattern behind operators under (\ref{eq: standard global symmetry}) and the polarization (\ref{eq: standard tbc}):
\begin{equation}
\begin{split}
	&\text{local operator from $\eta_{\text{D3}}$}: \text{D3-branes wrapping cone($\gamma_3^{(1)}$) },\\
	&\text{invertible $\mathbb{Z}_2^a$ line from $\mathcal{N}_{\text{D3}}$}:\text{D3-branes wrapping $\gamma_3^{(2)}$ ``at infinity".}
\end{split}
\end{equation}
This type of brane pattern falls in the general idea of branes ``at infinity'' as generalized symmetry operators introduced in \cite{Apruzzi:2022rei, GarciaEtxebarria:2022vzq, Heckman:2022muc}. The charged local operators and topological defect lines for the global symmetry $\mathbb{Z}_2^b$ and $\mathbb{Z}_2^c$ can be read similarly:
\begin{equation}
	\begin{split}
	&\text{local operator from $\eta_{\text{F1}}$}: \text{F1-strings  wrapping cone($\gamma_2$) },\\
	&\text{invertible $\mathbb{Z}_2^b$ line from $\mathcal{N}_{\text{NS5}}$}:\text{NS5-branes wrapping $\gamma_5$ ``at infinity",}\\
	&\text{local operator from $\eta_{\text{D1}}$}: \text{D1-strings wrapping cone($\gamma_2$) },\\
	&\text{invertible $\mathbb{Z}_2^c$ line from $\mathcal{N}_{\text{D5}}$}:\text{D5-branes wrapping $\gamma_5$ ``at infinity",}
\end{split}
\end{equation}

\subsection{Polarizations with non-invertible symmetries}
In the standard polarization $L=\mathbb{Z}^{\hat{a}}_2\oplus \mathbb{Z}^{\hat{b}}_2 \oplus \mathbb{Z}^{\hat{c}}_2$, there is a mixed anomaly for the global symmetry $G=\mathbb{Z}_2^a\times \mathbb{Z}^b_2\times\mathbb{Z}^c_2$, inherited from the Dijkgraaf-Witten twist in the 3D symmetry TFT:
\begin{equation}\label{eq: mixed anomaly}
	\pi \int_{M_3} a_1b_1c_1.
\end{equation}
According to \cite{Kaidi:2021xfk} (see also \cite{Kaidi:2023maf}), gauging two of the three $\mathbb{Z}_2$ symmetries, the left-over one will be promoted to a non-invertible symmetry. Let us take gauging $\mathbb{Z}_2^{a}\times \mathbb{Z}_2^b$ as an example. From the symmetry TFT perspective, this gauging process translates into changing the original Dirichlet boundary condition for $a_1$ and $b_1$ fields to Neumann boundary conditions. Their canonical conjugate $\hat{a}_1$ and $\hat{b}_1$ then pick Dirichlet boundary conditions accordingly. The resulting gapped boundary condition reads
\begin{equation}\label{eq: non-standard tbc}
   \hat{a}_1, \hat{b}_1, c_1~\text{Dirichlet}; ~a_1,b_1, \hat{c}_1~\text{Neumann},
\end{equation}
which picks a new polarization associated with the Lagrangian subgroup
\begin{equation}\label{eq: non-standard polar}
	L=\mathbb{Z}_2^a \times \mathbb{Z}_2^b \times \mathbb{Z}_2^{\hat{c}}.
\end{equation}

As we did in the ``standard'' polarization case, we can investigate the fate of various line operators in Table \ref{tab: branes behind lines} under the gapped condition (\ref{eq: non-standard tbc}) to investigate their roles in the resulting 2D absolute QFT. It is easy to see now 
\begin{equation}\label{eq: non-stand charged}
	\eta_{\text{D1}},\mathcal{N}_{\text{D3}}, \mathcal{N}_{\text{NS5}}
\end{equation}
are terminating on the gapped boundary, thus corresponding to local operators in the 2D QFT, while 
\begin{equation}\label{eq: non-stand top}
	\eta_{\text{D3}},\eta_{\text{F1}}, \mathcal{N}_{\text{D5}}
\end{equation}
can continue along the gapped boundary, thus corresponding to the topological defect line. Furthermore, the Neumann boundary condition for $a_1,b_1$ and $\hat{c}_1$ preserves the non-invertible property for the $\mathcal{N}_{\text{D5}}$ line, so it still reads
\begin{equation}\label{eq: non-inver D5 line}
	\mathcal{N}_{\text{D5}}= \int\mathcal{D}\hat{\phi}_0 \mathcal{D}\phi_0 e^{ \pi i \int_{M_1} \hat{c}_1 }  e^{ \pi i \int_{M_1} \hat{\phi}_0\delta \phi_0+\phi_0 a_1-\hat{\phi}_0 b_1 }.
\end{equation}

Therefore, based on its fusion rule (\ref{eq:fusion of D5 operator}), we conclude the global symmetry for the polarization (\ref{eq: non-standard polar}) is 
\begin{equation}
	G=\mathbb{Z}_2^{\hat{a}}\times \mathbb{Z}_2^{\hat{b}}~\text{Tambara-Yamagami Categorical Symmetry}.
\end{equation}
The brane pattern for this global symmetry can be built by wrapping branes in (\ref{eq: non-stand charged}) terminating ``at infinity'' as charged operators while wrapping branes in (\ref{eq: non-stand top}) ``at infinity'' as topological defect lines.

Field theoretically, one would follow the step in \cite{Kaidi:2021xfk} to compute what would be a non-invertible TFT promote of the invertible $\mathbb{Z}_2^c$ line after gauging with the presence of the mixed anomaly (\ref{eq: mixed anomaly}). The result will perfectly coincide with (\ref{eq: non-inver D5 line})\footnote{We thank Ho Tat Lam for valuable discussions on this point.}. The punchline of our top-down approach is that the non-invertible line directly comes from the D5-brane worldvolume action, as we computed in Section 2. Its (non-)invertible property in absolute QFTs before/after gauging simply results from changing polarizations, which translates into different brane patterns ``at infinity''.

We remark that the non-invertible symmetry in this context is known as the non-intrinsic one \cite{Kaidi:2022uux}. This is because it is related to an invertible symmetry via changing polarizations\footnote{For discussion on intrinsic vs. non-intrinsic non-invertible symmetries from higher-dimensional perspective, we refer the reader to \cite{Heckman:2022xgu, Bashmakov:2022uek}}.

We conclude this subsection by presenting three polarizations that enjoy non-invertible symmetries and their comparison with the ``standard'' polarization in Table \ref{tab: polarizations and symmetries}.
\begin{table}
    \centering
    \begin{tabular}{|c|c|c|c|}
\hline
Polarization $L$ & Global Symmetry  & Charged Operators & Symmetry Lines \\
     \hline
      $\mathbb{Z}_2^{\hat{a}}\times \mathbb{Z}_2^{\hat{b}} \times \mathbb{Z}^{\hat{c}}_2$ & $\mathbb{Z}_2^{a}\times \mathbb{Z}^b_2 \times \mathbb{Z}^c_2$ & $\eta_{\text{D3}},\eta_{\text{F1}}, \eta_{\text{D1}}$ & $\mathcal{N}_{\text{D3}}, \mathcal{N}_{\text{NS5}}, \mathcal{N}_{\text{D5}}$\\
      \hline
      $\mathbb{Z}_2^{a}\times \mathbb{Z}_2^{b} \times \mathbb{Z}^{\hat{c}}_2$ & $\mathbb{Z}_2^{\hat{a}}\times \mathbb{Z}^{\hat{b}}_2$ TY category & $\mathcal{N}_{\text{D3}}, \mathcal{N}_{\text{NS5}}, \eta_{\text{D1}}$ & $\eta_{\text{D3}},\eta_{\text{F1}}, \mathcal{N}_{\text{D5}}$\\      
      \hline
      $\mathbb{Z}_2^{a}\times \mathbb{Z}_2^{\hat{b}} \times \mathbb{Z}^{c}_2$ & $\mathbb{Z}_2^{\hat{a}}\times \mathbb{Z}^{\hat{c}}_2$ TY category & $\mathcal{N}_{\text{D3}},  \eta_{\text{F1}}, \mathcal{N}_{\text{D5}}$ & $\eta_{\text{D3}},\eta_{\text{D1}}, \mathcal{N}_{\text{NS5}}$\\
      \hline
    $\mathbb{Z}_2^{\hat{a}}\times \mathbb{Z}_2^{b} \times \mathbb{Z}^{c}_2$ & $\mathbb{Z}_2^{\hat{b}}\times \mathbb{Z}^{\hat{c}}_2$ TY category & $\eta_{\text{D3}},  \mathcal{N}_{\text{NS5}},\mathcal{N}_{\text{D5}}$ & $\eta_{\text{F1}},\eta_{\text{D1}}, \mathcal{N}_{\text{D3}}$\\
      \hline
\end{tabular}
    \caption{A ``Standard'' polarization with the $(\mathbb{Z}_2)^3$ invertible symmetry, as well as three polarizations with the $\mathbb{Z}_2\times \mathbb{Z}_2$ Tambara-Yamagami (TY) categorical symmetry. The concrete torsional cycles wrapped by branes for various $\eta$ and $\mathcal{N}$ operators can be found in Table \ref{tab: branes behind lines}. The charge operators are built by branes terminating at the asymptotic boundary $Y^{2,0}(\mathbb{P}^1\times \mathbb{P}^1)$, while the symmetry lines are built by branes ``at infinity'' along the asymptotic boundary.}
    \label{tab: polarizations and symmetries}
\end{table}

\subsection{Action of non-invertible lines and the Hanany-Witten transition}
One salient property of the non-invertible symmetry defect is its action on the charged operator (see, e.g., \cite{Komargodski:2020mxz, Choi:2021kmx}). Consider the polarization (\ref{eq: non-standard polar}) with the non-invertible line $\mathcal{N}_{\text{D5}}$. Moving this line past the local operator charged under $\mathbb{Z}_2^{\hat{a}}$ (resp. $\mathbb{Z}_2^{\hat{b}}$) will make the charged operator non-genuine and attach to the topological $\mathbb{Z}_2^{\hat{a}}$ (resp. $\mathbb{Z}_2^{\hat{b}}$) topological line. Namely, it belongs to the defect Hilbert space of the line it attached \cite{Chang:2018iay}.

\begin{figure}[H]
	\centering
	\includegraphics[width=15cm]{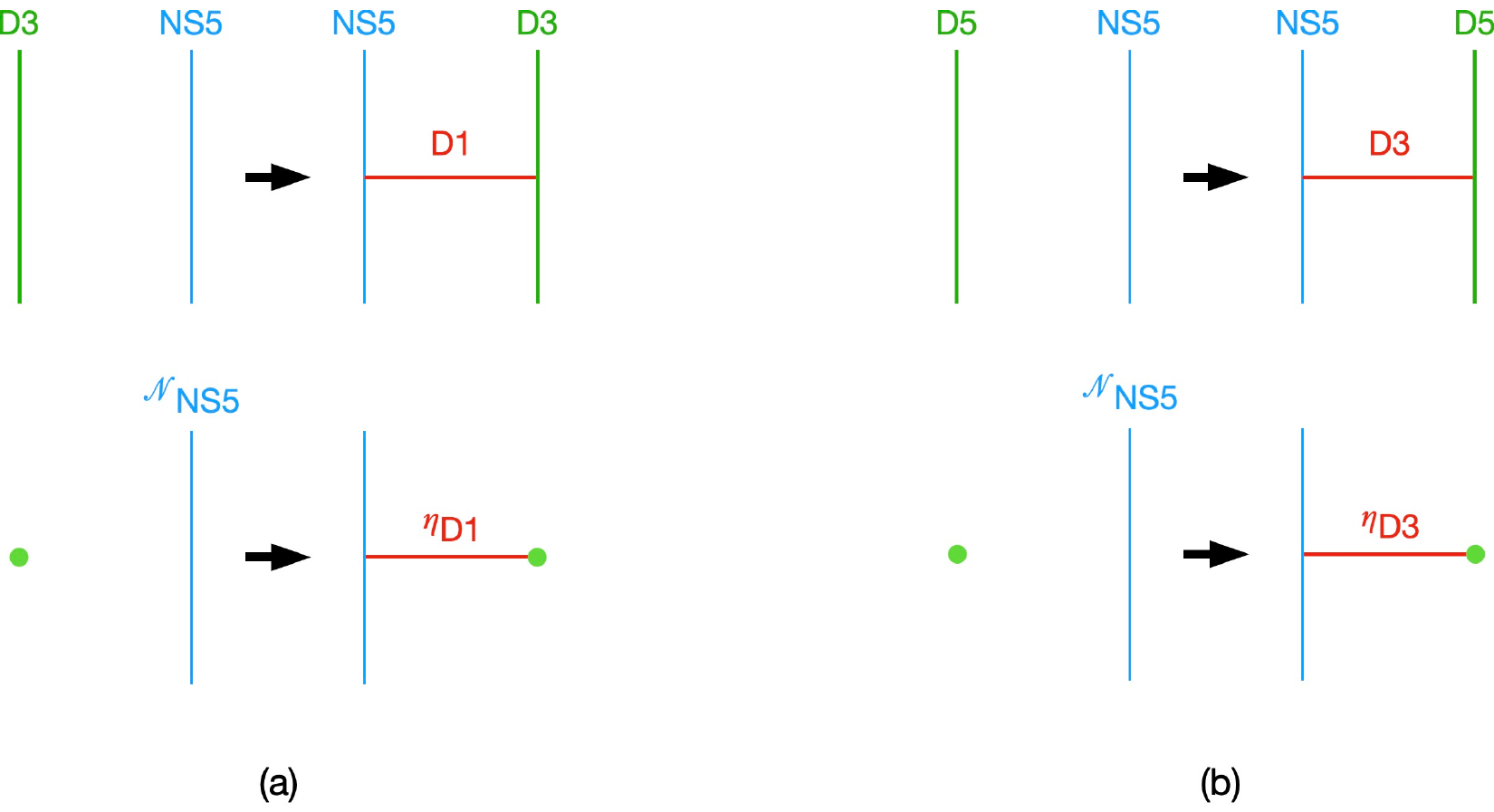}
	\caption{Bottom: Action of non-invertible defect line $\mathcal{N}_{\text{NS5}}$ (in the polarization $L=\mathbb{Z}_{2}^{a}\times \mathbb{Z}_2^{\hat{b}}\times \mathbb{Z}_2^c$) on the local operators charged under the invertible symmetry $\mathbb{Z}_2^{\hat{a}}$ and $\mathbb{Z}_2^{\hat{c}}$. The corresponding topological defect lines are generated under this action. Top: This non-trivial action enjoys a string theory origin as the Hanany-Witten transition, where the created branes wrapping cycles ``at infinity'' perfectly serve as the topological defect lines.}
	\label{fig: HWtransition}
\end{figure}

This non-trivial action enjoys an elegant string theory origin as the Hanany-Witten transition \cite{Hanany:1996ie}. Note that the charged operator under $\mathbb{Z}_2^{\hat{a}}$ (resp. $\mathbb{Z}_2^{\hat{b}}$) comes from the D3-brane wrapping on cone$(\gamma_3^{(2)})$ (resp. NS5-brane wrapping on cone($\gamma_5$)). When the D5-brane generating the non-invertible $\mathcal{N}_{\text{D5}}$ passes through the above D3-brane (resp. NS5-brane), a F1 string wrapping $\gamma_1$ (resp. D3-brane wrapping $\gamma_3^{(1)}$) is generated connecting them. What object is generated by the F1-string wrapping $\gamma_1$ (resp. D3-brane wrapping $\gamma_3^{(1)}$)? It is exactly the topological defect line $\eta_{\text{F1}}$ (resp. $\eta_{\text{D3}}$) for the $\mathbb{Z}_2^{\hat{a}}$ (resp. $\mathbb{Z}_2^{\hat{b}}$) symmetry (see Table \ref{tab: branes behind lines} and \ref{tab: polarizations and symmetries})! See Figure \ref{fig: HWtransition} for an illustration of how the Hanany-Witten transition translates into non-trivial transitions of charged operators in the polarization, e.g.,  $L=\mathbb{Z}_{2}^{a}\times \mathbb{Z}_2^{\hat{b}}\times \mathbb{Z}_2^c$. 

Similar Hanany-Witten transition origin for the non-trivial action of non-invertible defect has been observed in 4D QFTs \cite{Apruzzi:2022rei, Heckman:2022xgu}. Still, to our knowledge, in the context of 2D QFTs, this is the first time the Hanany-Witten interpretation of this action appears in the literature. It is, therefore, natural to conjecture that for non-invertible symmetries with brane origins\footnote{There are also cases where this non-trivial action is engineered from purely geometric setup, see, e.g., \cite{Lawrie:2023tdz}.}, this correspondence holds for diverse dimensions \footnote{Recently, this non-trivial action is formulated in the language of the higher representation and generalized charges in reference \cite{Apruzzi:2023uma}.}.

\section{Future Directions}
Our work suggests various natural directions for future investigation. Some obvious and interesting directions include:

\begin{itemize}
    \item As we pointed out at the beginning of this note, there is so far an infinite family of 2D gauge theories arising from D1-branes probing toric Calabi-Yau 4-folds. Investigating their symmetry TFTs, non-invertible, and other generalized symmetries would be interesting. We will explore this subject in \cite{Franco:2023toapp}. 
    \item There are four fusion categories associated with the same $\mathbb{Z}_2\times \mathbb{Z}_2$ TY fusion rule, but distinguished by their associators or F-symbols \cite{TAMBARA1998692}. Three of those are given by the representations as Rep$(D_4)$, Rep$(Q_8)$ and Rep$(\mathcal{H}_8)$\footnote{$\mathcal{H}_8$ is the 8-dimensional Kac-Paljutkin Hopf algebra \cite{kac1966finite}}. Identifying which corresponds to the categorical symmetry we derived in this note would be interesting. This example may shed new light on a general question: Given a categorical symmetry with certain fusion rules from string theory, is there any top-down approach to its F-symbols or (generalized) Frobenius-Schur indicators? We expect this 2D example to be a nice starting point to answer this question in diverse dimensions.
    \item Based on the above direction, it would also be interesting to investigate anomalies and gauging of non-invertible symmetries from string theory perspectives, following the purely field-theoretic consideration \cite{Fuchs:2002cm, Bhardwaj:2017xup, Zhang:2023wlu, Perez-Lona:2023djo, Diatlyk:2023fwf}.
\end{itemize}

\section*{Acknowledgements}
XY thanks S. Franco, I. Garcia Etxebarria, J. J. Heckman, M. H{\"u}bner, H. T. Lam, S. Schafer-Nameki, S.-H. Shao, E. Sharpe, E. Torres, Y. Wang, and H. Y. Zhang for helpful discussions. XY also thanks H. T. Lam for his collaboration in the early phase of this project. XY thanks the UPenn Theory Group for their hospitality during part of this work. XY would like to thank the 2023 Simons Physics Summer Workshop for the hospitality during part of this work. The work of XY was supported in part by the NYU James Arthur Graduate Associate Fellowship. The work of XY is supported by NSF grant PHY-2014086.

\appendix

\bibliographystyle{utphys}
\bibliography{2dnoninver}

\end{document}